\documentclass{ifacconf}
\makeatletter

\let\AND\@undefined
\usepackage{amsmath,amssymb}
\usepackage{algorithm,algorithmic}
\usepackage{refcount}
\usepackage{pdfpages}



\usepackage{color}
\usepackage{graphicx}      
\usepackage{natbib}        
\usepackage{setspace}
\usepackage{xspace}
\usepackage{subcaption}

\newcommand{\s}{x}
\newcommand{\T}{T}
\newcommand{\out}{y}

\newcommand{\Exp}[1]{\operatorname{\mathbb{E}}\left[#1\right]}
\newcommand{\N}{\mathcal{N}}
\newcommand{\CPFAS}{CPF-AS\xspace}

\newcommand{\W}{W}
\newcommand{\anc}{a}
\newcommand{\Prob}[1]{\mathbb{P}\left(#1\right)}
\newcommand{\Kernel}{\mathcal{K}}
\newcommand{\kmcmc}{k}
\newcommand{\Kmcmc}{K}
\newcommand{\statp}{\pi}
\newcommand{\Expb}[2]{\operatorname{\mathbb{E}}_{#1}\left[#2\right]}

\newcommand{\FFBSi}{\textrm{FFBSi}\xspace}
\newcommand{\ffbsi}{\textrm{FFBSi}\xspace}
\newcommand{\mcmc}{\textrm{MCMC}\xspace}

\newcommand{\SSM}{\textrm{SSM}\xspace}

\newcommand{\proposal}{q}
\newcommand{\nx}{{n_x}}

\newcommand{\w}{w}	

\begin{document}
\begin{frontmatter}

\title{Nonlinear State Space Smoothing Using the Conditional Particle Filter\thanksref{footnoteinfo}}

\thanks[footnoteinfo]{This work was supported by the project \emph{Probabilistic modelling of dynamical systems} (Contract number: 621-2013-5524) and CADICS, a Linnaeus Center, both funded by the Swedish Research Council (VR).}

\author[UU]{Andreas Svensson} 
\author[UU]{Thomas B. Sch\"{o}n}
\author[LiU]{Manon Kok} 

\address[UU]{Department of Information Technology, Uppsala University, Sweden (e-mail: \{andreas.svensson, thomas.schon\}@it.uu.se)}
\address[LiU]{Division of Automatic Control, Link\"{o}ping University, Sweden (e-mail: manon.kok@liu.se)}

\begin{abstract}                
To estimate the smoothing distribution in a nonlinear state space model, we apply the conditional particle filter with ancestor sampling. This gives an iterative algorithm in a Markov chain Monte Carlo fashion, with asymptotic convergence results. The computational complexity is analyzed, and our proposed algorithm is successfully applied to the challenging problem of sensor fusion between ultrawideband and accelerometer/gyroscope measurements for indoor positioning. It appears to be a competitive alternative to existing nonlinear smoothing algorithms, in particular the forward filtering-backward simulation smoother.
\end{abstract}
%
%
\end{frontmatter}

\section{Introduction}
Consider the (time-varying, nonlinear, non-Gaussian) state space model (\SSM)
\begin{subequations}
  \label{eq:ss}
  \begin{align}
    \s_{t+1}\mid \s_t & \sim f_{t}(\s_{t+1}|\s_{t}), \\
    \out_t\mid\s_t & \sim g_{t}(\out_t|\s_t),
  \end{align}
\end{subequations}
with $\s_1 \sim \mu(\s_1)$. We use a probabilistic notation, with $\sim$ meaning distributed according to. The index variable $t = 1, 2, \dots, T$ is referred to as time. The variable $\s_t \in \mathbb{R}^{\nx}$ is referred to as state, and an exogenous input $u_t$ is possible to include in $f_t$ and $g_t$. To ease the notation, the possible time dependence of $f$ and $g$ will be suppressed.

\begin{figure}[t!]
	\vspace{-0.8em}
	\begin{center}
		\includegraphics[width=0.96\columnwidth]{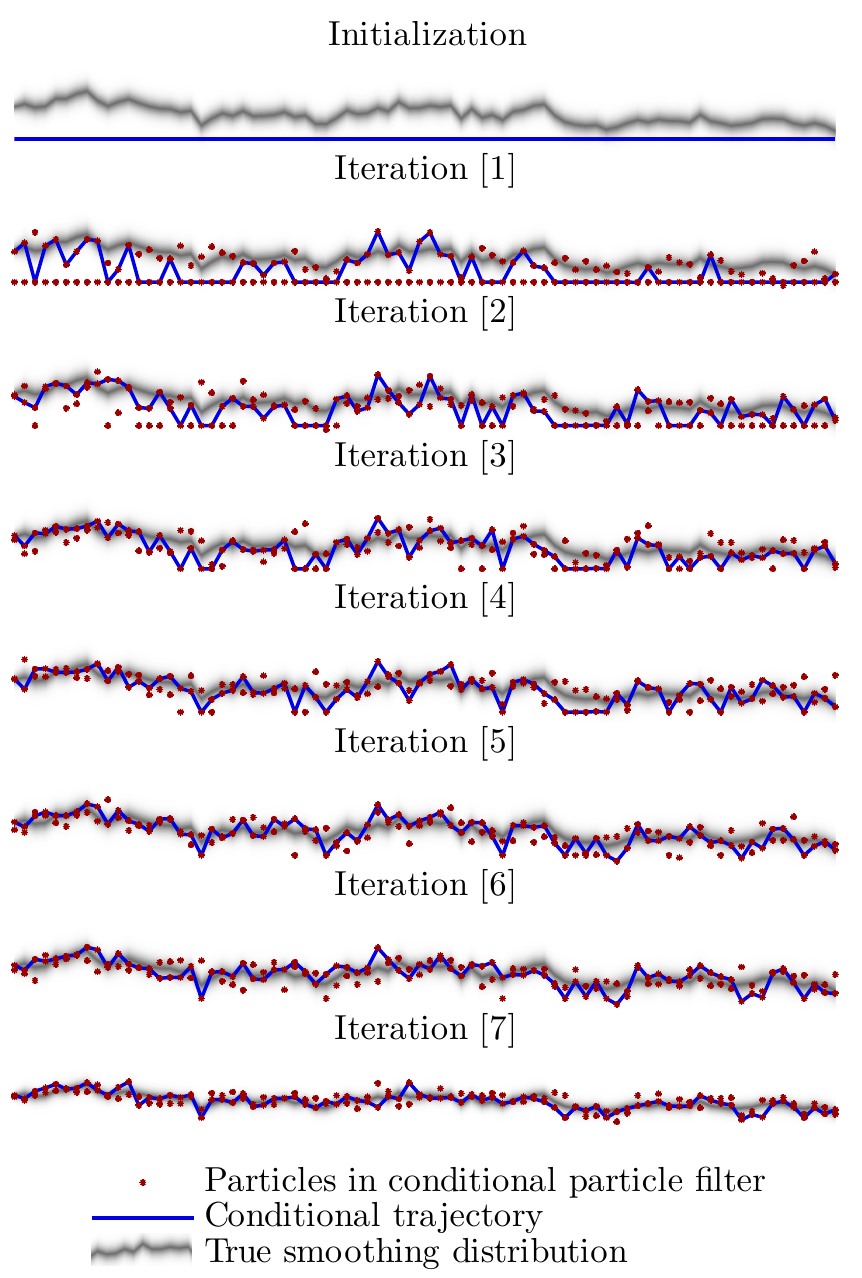}
	\end{center}
	\caption{\small How to use the conditional particle filter to sample from the smoothing distribution $p(x_{1:T}|y_{1:T})$. The distribution $p(x_{1:T}|y_{1:T})$ is shown in gray, with time $t$ at the horizontal axis and state $\s$ on the vertical axis. The conditional particle filter (with only 2 particles) is run iteratively, starting with the its arbitrary initialized trajectory (blue line) in the top plot, eventually converging to provide samples from the smoothing distribution.}
	\label{fig:intro}
\end{figure}

For some applications, e.g.,  system identification, the distribution of the states for given model and measurements,
\begin{align}
p(\s_{1:T}|\out_{1:T}),\label{eq:jsd}
\end{align}
is of interest. We will refer to \eqref{eq:jsd} as the \emph{smoothing distribution}. The smoothing distribution is not available on closed form for the general model~\eqref{eq:ss}, and approximations are necessary. In this paper, we will present a method generating Monte Carlo samples, \emph{particles}, from the smoothing distribution, akin to a particle filter. The idea is to iterate a \emph{conditional particle filter}, which generates samples from the smoothing distribution after sufficiently many iterations, as illustrated in Figure~\ref{fig:intro}.

An overview of existing particle smoothers addressing the problem of generating samples from \eqref{eq:jsd} is provided by \citet{lindsten2013backward}. In this work we will in particular compare and relate our developments to the so-called forward filtering-backward simulation (\ffbsi) smoother introduced by \citet{douc2011}. The work by \citet{kitagawa1996monte} and \citet{briers2010smoothing} are both of interest in that they approach a similar problem using particle filters, the latter also taking inspiration from the two-filter formula. The work of \citet{PillonettiB2008} is closely related in that they also employ a Markov chain Monte Carlo (\mcmc) construction, but they consider the special case of Gaussian noise ($f$ and $g$ can still be nonlinear though).

An alternative to using samples to represent \eqref{eq:jsd} is to solve a linearized version of the problem, by combining the extended Kalman filter \citep{smith1962application,schmidt1966application} with the RTS-smoother \citep{rauch1965maximum} to solve the linearized problem analytically. The work of \cite{sarkka2008unscented} on the unscented RTS-smoother is along the same line.

It should also be possible to generalise the ideas presented in this paper to probabilistic graphical models along the lines of the work by \cite{NLS:2014}.


Source code for simulated examples are available via the first author's homepage, 
and details on the second simulated example and the indoor positioning problem are available in a technical report \citep{techreport}.

\section{Particle methods}\label{sec:particlemethods}

We assume the reader has some basic familiarity with particle filters (see, e.g., \cite{doucet2009tutorial} for an introduction), but to set the notation we will start by a brief summary of particle filters and particle smoothers.

\subsection{Particle filters}
In the search for a numerical approximation to $p(\s_{1:T}|\out_{1:T})$, the following factorization is useful
\begin{equation}
p(\s_{1:T},\out_{1:T}) = \mu(\s_1)\prod_{t=1}^Tg(\out_t|\s_t)\prod_{t=1}^{T-1}f(x_{t+1}|x_t), \label{eq:ssfact}
\end{equation}
as it allows for the following recursion to be derived (using Bayes' rule and $p(\out_{1:T}) = p(\out_1)\prod_{t=1}^{T-1}p(\out_{t+1}|y_{1:t})$)
\begin{equation}
p(\s_{1:t}|\out_{1:t}) 
=\underbrace{\frac{g(\out_t|\s_t)f(\s_t|\s_{t-1})}{p(\out_t|\out_{1:t-1})}}_{(\ref{eq:ffilteq}\star)}p(\s_{1:t-1}|\out_{1:t-1}).\label{eq:ffilteq}
\end{equation}
This factorization can be used to motivate the particle filter. Starting with a particle (Monte Carlo) approximation of $p(x_1|y_1)$ as $N$ particles, (\ref{eq:ffilteq}$\star$) can be applied to obtain a particle approximation of $p(\s_{1:2}|\out_{1:2})$. Repeating this $T-1$ times, a particle approximation of $p(\s_{1:T}|\out_{1:T})$ as $N$ weighted particles $\{\s_{1:T}^i,\w_T^i\}_{i=1}^N$ is found. This is detailed in Algorithm~\ref{alg:pf}, the \emph{particle filter}, a Sequential Monte Carlo method. 

The notation used in Algorithm~\ref{alg:pf} is 
\begin{subequations}
\begin{align}
\W_{1}(\s_1) &\triangleq g(\out_1|\s_1)\mu(\s_1)/q_{1}(\s_1|y_1)\\
\W(\s_{1:t}) &\triangleq 
g(\out_t|\s_t)f(\s_t|\s_{t-1})/\proposal(\s_t|\s_{t-1},\out_t).\label{eq:W}
\end{align}
\end{subequations}
Here, $\proposal$ denotes the proposal distribution, which is used to propagate the particles from time $t$ to time $t+1$. If the proposal $\proposal$ is chosen as $f$, then \eqref{eq:W} simplifies to $\W(x_{1:t}) = g(y_t|x_t)$ resulting in the so-called bootstrap particle filter.

The main steps in Algorithm~\ref{alg:pf}, namely 4, 5 and 7 are often referred to as \emph{resampling}, \emph{propagation} and \emph{weighting}, respectively. Step 6 is merely bookkeeping. In Algorithm~\ref{alg:pf}, a notation using \emph{ancestor indices} $\anc_t^i$ has been used for the resampling step, to prepare for the expansion to conditional particle filter with ancestor sampling.

\begin{algorithm}[t]
	\caption{Particle Filter}
	\label{alg:pf}
	\begin{algorithmic}[1]
		\ENSURE A weighted particle system $\{\s_{1:t}^i,\w_t^i\}_{i=1}^N$\\ approximating $p(\s_{1:t}|\out_{1:t})$ for $t = 1, 2, \dots, T$.
		\STATE Draw $\s_1^i \sim q_{1}(\s_i)$ for $i = 1, \dots N$.
		\STATE Set $w_1^i = \W_{1}(\s_1^i)$ for $i = 1, \dots N$.
		\FOR{$t = 2, \dots, T$}
		\STATE\label{alg:pf:res} Draw $\anc_t^i$ with $\Prob{\anc_t^i=j} \propto w_{t-1}^j$ for $i = 1, \dots N$.
		\STATE\label{alg:pf:prop} Draw $\s_t^i \sim q_t(\s_t|\s_{t-1}^{\anc_t^i},\out_t)$ for $i = 1, \dots N$.
		\STATE\label{alg:pf:book} Set $\s_{1:t}^i = \{x_{1:t-1}^{\anc_t^i},x_t^i\}$ for $i = 1, \dots N$.
		\STATE\label{alg:pf:weight} Set $\w_{t}^i = \W(\s_{1:t}^i)$ for $i = 1, \dots N$.
		\ENDFOR
	\end{algorithmic}
\end{algorithm}

\begin{figure}[t]
  \begin{center}
    \includegraphics[width=0.5\textwidth]{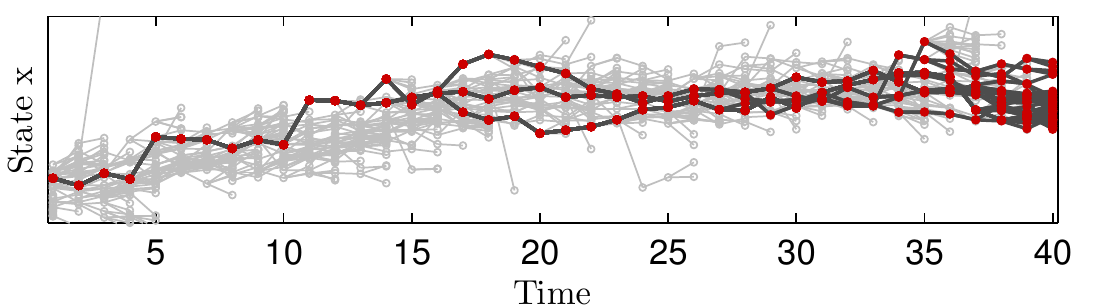}
  \end{center}
  \caption{\small Path degeneracy: The particles in a particle filter are shown as dots, propagated as the lines indicate. The red dots are particles that have `survived' the resampling steps, whereas the grey dots have not `survived' the resampling steps. All trajectories $\s_{1:40}^i$ have the part $\s_{1:13}$ in common. This phenomenon occurs in particle filters and is the reason why a particle filter does not provide a good numerical approximation of $p(\s_{1:T}|\out_{1:T})$ for a finite $N$.}
  \label{fig:path_degeneracy}
\end{figure}

In theory, a particle filter directly gives a numerical approximation of $p(\s_{1:T}|\out_{1:T})$. However, in practice with a finite $N$, the approximation tends to be rather poor (unless $\T$ is very small), as it typically suffers from \emph{path degeneracy} as illustrated in Figure~\ref{fig:path_degeneracy}.

\subsection{Forward -- backward particle smoothers}\label{sec:ffbsi}

A natural way to find the smoothing distribution for SSMs is to first apply a (forward) filter, and then add a backward pass, adjusting for the `new' information about the state $\s_t$ at time $t$ obtained from the later measurements $\out_{t+1}, \dots, \out_T$. Such an example is the RTS smoother for the linear Gaussian case \citep{rauch1965maximum}, but also the more recent particle-based \FFBSi algorithm, see, e.g., \cite{lindsten2013backward} for a recent overview.

The algorithm for $\FFBSi$ is not repeated here, but we note that the it relies on the two step
\begin{enumerate}
\item\label{ffbsi:1} A particle filter with $N$ particles.
\item A backward simulation drawing $M$ (uncorrelated) samples from $p(\s_{1:T}|y_{1:T})$ using the $N$ particles from Step~(\ref{ffbsi:1}).
\end{enumerate}
The computational complexity of \FFBSi is basically $\mathcal{O}(NM)$, although some improvements can be achieved, see \cite[Section 3.3]{lindsten2013backward}.

To prepare for the upcoming discussions on convergence, let us briefly comment on the convergence properties of \ffbsi.

How well can a function $h(x_{1:T})$ be approximated as $\bar{h}$ using samples from \FFBSi ? Let $\bar{h}_{\text{\FFBSi}}^N = \frac{1}{N}\sum_{i = 1}^N h(\s_{1:T}^i)$ denote an approximation of $h(\s_{1:T})$ based on $M = N$ backward trajectories. Under some fairly mild assumptions, it has been shown \cite[Corollary 9]{douc2011} that there exists a $\sigma_{\text{\FFBSi}} < \infty$ such that
\begin{align}
\sqrt{N}\left(\bar{h}_{\text{\FFBSi}}^{N}-\Exp{h(\s_{1:T})|\out_{1:T}}\right)
\end{align}
converges weakly to $\mathcal{N}(0,\sigma_{\text{\FFBSi}}^2)$.

To summarize, the convergence rate for \FFBSi is $\sqrt{N}$, subject to a computational complexity of $\mathcal{O}(N^2)$.

\section{Smoothing using the Conditional Particle Filter}\label{sec:pmcmc}

The smoothing methodology discussed in Section~\ref{sec:ffbsi} builds on a forward-backward strategy. The MCMC idea offers a fundamentally different way to construct a smoother, \emph{without} explicitly running a backward pass, but \emph{iteratively} running a so-called \emph{conditional particle filter} as illustrated in Figure~\ref{fig:intro}. As we will see, this opens up for a reduced computational complexity. The origin of the method dates back to the introduction of the PMCMC methods by \cite{andrieu2010particle}, with important recent contributions from \cite{lindsten2014particle}.

First, the conditional particle filter with ancestor sampling (\CPFAS) will be introduced (Section \ref{sec:cpfas}), followed by a brief introduction to MCMC (Section \ref{sec:mcmc}), and they will in the next step (Section~\ref{sec:pmcmcm_smoothing}) be combined to form a particle smoother. The convergence properties and the computational complexity of the smoother are then examined in Section~\ref{sec:cpfas_convergence} and Section~\ref{sec:cpfas_cc}, respectively.

\subsection{Conditional particle filter with ancestor sampling}\label{sec:cpfas}
The \CPFAS is thoroughly described by \cite{lindsten2014particle}, and here presented as Algorithm~\ref{alg:cpfas}. The \CPFAS is similar to a regular particle filter, Algorithm~\ref{alg:pf}, in many aspects, but with one particle trajectory $\s_{1:T}[k]$ specified a priori (trajectory number $N$ in Algorithm~\ref{alg:cpfas}).

The \CPFAS generates $N$ weighted particle trajectories $\{\s_{1:T}^i,\w_T^i\}_{i=1}^N$. With the original formulation of the conditional particle filter in \cite{andrieu2010particle}, one of these trajectories is predestined to be $\s_{1:T}[k]$. Extending this with ancestor sampling, the \CPFAS is obtained and the resulting trajectories $\{\s_{1:T}^i,\w_T^i\}_{i=1}^N$ are still influenced by $\s_{1:T}[k]$, but in a somewhat more involved way, as the conditional trajectory may be `partly' replaced by a new trajectory; see Algorithm~\ref{alg:cpfas} for details. 

By sampling one of the trajectories $\s_{1:T}[k+1] = \s_{1:T}^J$ obtained from the \CPFAS with $\Prob{i=J}\propto\w_T^i$, the \CPFAS can be seen as a procedure to stochastically map $\s_{1:T}[k]$ onto another trajectory $\s_{1:T}[k+1]$.

\begin{algorithm}
\caption{Conditional particle filter with ancestor sampling (\CPFAS)}
\label{alg:cpfas}
\begin{algorithmic}[1]
\REQUIRE Trajectory $\s_{1:T}[k]$
\ENSURE Trajectory $\s_{1:T}[k+1]$
\STATE Draw $\s_1^i \sim q_{1}(\s_1^i)$ for $i = 1,\dots,N-1$.
\STATE Set $\s_{1}^N = \s_{1}[k]$.
\STATE Set $w_1^i = \W_{1}(\s_1^i)$ for $i = 1, \dots, N$.
\FOR{$t = 2, \dots, T$}
\STATE Draw $\anc_t^i$ with $\Prob{\anc_t^i=j} \propto w_{t-1}^j$ for $i = 1, \dots, N-1$.
\STATE Draw $\s_t^i \sim \proposal(\s_t|\s_{t-1}^{\anc_t^i},\out_t)$ for $i = 1, \dots, N-1$.
\STATE \label{alg:cpfas:Nprop} Set $\s_t^N = \s_t[k]$.
\STATE \label{alg:cpfas:Nanc} Draw $\anc_t^N$ with $\Prob{\anc_t^N = j} \propto w_{t-1}^jf(\s_t^N|\s_{t-1}^j)$.\label{alg:cpfas:ancs}
\STATE Set $\s_{1:t}^i = \{x_{1:t-1}^{\anc_t^i},x_t^i\}$ for $i = 1, \dots, N$.
\STATE \label{alg:cpfas:weight} Set $\w_{t}^i = \W(\s_{1:t}^i)$ for $i = 1, \dots, N$.
\ENDFOR
\STATE Draw $J$ with $\Prob{i=J}\propto \w_T^i$ and set $\s_{1:T}[k+1] = \s_{1:T}^J$.
\end{algorithmic}
\end{algorithm}


A Rao-Blackwellized formulation of the \CPFAS for mixed linear/nonlinear models is also possible, see \cite{svensson2014rbpsaem} for details.

\subsection{Markov chain Monte Carlo}\label{sec:mcmc}

MCMC offers a strategy for sampling from a complicated probability distribution $\pi$ on the space $\mathcal{Z}$, using an iterative scheme.

A Markov chain on $\mathcal{Z}$ is a sequence of the random variables $\{\zeta[1], \zeta[2], \zeta[3], \dots\}$, $\zeta[\kmcmc] \in \mathcal{Z}$. The chain is defined by a \emph{kernel} $\Kernel$, stochastically mapping one element $\zeta[\kmcmc]$ onto another element $\zeta[\kmcmc+1]$. That is, the distribution of the random variable $\zeta[\kmcmc]$ depends on the previous element as $\zeta[\kmcmc+1] \sim \Kernel(\cdot | \zeta[\kmcmc]))$. 

If the kernel $\Kernel$ is \emph{ergodic} with a unique stationary distribution $\statp$, the marginal distribution of the chain will approach $\statp$ in the limit. Let $\zeta[0]$ be an arbitrary initial state with $\statp( \zeta[0]) > 0$, then by the ergodic theorem \citep{Robert2004mcs}
\begin{align}
\frac{1}{K}\sum_{\kmcmc=1}^\Kmcmc h(\zeta[\kmcmc]) \rightarrow \Expb{\statp}{ h(\zeta) }, \label{eq:Kmcmc}
\end{align}
as $\Kmcmc \rightarrow \infty$ for any function $h: \{\mathbb{R}^{n}\}^T \mapsto \mathbb{R}$, with $\Expb{\statp}{\cdot}$ denoting expectation w.r.t. $\zeta$ under the distribution $\statp$.

That is, for sufficient large $\kmcmc$, the realization of $\{\zeta[\kmcmc], \zeta[\kmcmc+1], \dots\}$ is (possibly correlated) samples from $\statp$. This summarizes the idea of the MCMC methodology; if $\statp$ is of interest, construct a kernel $\Kernel$ with stationary distribution $\statp$ and simulate a Markov chain to obtain samples of $\statp$.

Note that any \emph{finite} realization of the chain $\{\zeta[1], \dots, \zeta[\Kmcmc]\}$ may be an arbitrarily bad approximation of $\statp$. This typically depends on the initialization $\zeta[0]$ and on how well the kernel $\Kernel$ manages to explore $\mathcal{Z}$, referred to as the \emph{mixing}.

\subsection{Smoothing using MCMC}\label{sec:pmcmcm_smoothing}

Take the general space $\mathcal{Z}$ as the more concrete space $\{\mathbb{R}^{n_x}\}^\T$ (where $\s_{1:T}$ lives). Note that \CPFAS in Algorithm~\ref{alg:cpfas} maps one element in $\{\mathbb{R}^\nx\}^\T$ onto another element in $\{\mathbb{R}^\nx\}^\T$, and can therefore be interpreted as an MCMC kernel. The unique stationary distribution for \CPFAS is $p(\s_{1:T}|\out_{1:T})$ (which is far from obvious, but shown by \cite{lindsten2014particle}).  Now, by constructing a Markov chain, Algorithm~\ref{alg:mcmc_smoother} is obtained, generating samples from the distribution $p(\s_{1:T}|\out_{1:T})$ (i.e., a smoother).

\begin{algorithm}[t]
\caption{MCMC smoother}
\label{alg:mcmc_smoother}
\begin{algorithmic}[1]
\REQUIRE $\s_{1:T}[0]$ (Initial (arbitrary) state trajectory)
\ENSURE $\s_{1:T}[1], \dots, \s_{1:T}[K]$ ($K$ samples from the Markov chain)
\FOR{$k = 1, \dots, K$} 
\STATE Run the \CPFAS (Algorithm~\ref{alg:cpfas}) conditional on $\s_{1:T}[k-1]$ to obtain $\s_{1:T}[k]$.
\ENDFOR
\end{algorithmic}
\end{algorithm}

An illustration of Algorithm~\ref{alg:mcmc_smoother} was provided already by Figure \ref{fig:intro}; The initial trajectory is obviously not a sample from $\statp = p(\s_{1:T}|\out_{1:T})$, and artifacts from the initializations appear to be present also in iteration $[1]$, $[2]$, and possibly $[3]$. However, iterations $[5], [6], [7]$ appear to be (correlated) samples from the distribution $\statp$, which is what was sought.

\subsection{Convergence}\label{sec:cpfas_convergence}
The convergence analysis of Algorithm~\ref{alg:mcmc_smoother} can, similar to the \FFBSi in Section~\ref{sec:ffbsi}, be posed as the question of how well $h(\s_{1:T})$ can be approximated by $\bar{h}^{\Kmcmc}_{\text{\text{\CPFAS}}} = \frac{1}{K}\sum_{\kmcmc = 1}^\Kmcmc h(\s_{1:T}[\kmcmc])$, where $\s_{1:T}[\kmcmc]$ comes from Algorithm~\ref{alg:mcmc_smoother}. Before stating the theorem, let us make the following two rather technical assumptions
\begin{itemize}
\item[A1.] The proposal $\proposal$ is designed such that given any $x_{t-1}$ with non-zero probability (given the measurements $\out_{1:t-1}$), any $x_t$ with non-zero probability (given $\out_{1:t}$) should be contained in the support of $q$.
\item[A2.] There exists a constant $\kappa < \infty$ such that $\|W\|_\infty<\kappa$.
\end{itemize}
 
\begin{thm}[Convergence for Algorithm~\ref{alg:mcmc_smoother}]\label{thm:conv_rate_cpfas}
Under the assumptions A1 and A2, for any number of particles $N > 1$, and for any bounded function $h:~\{\mathbb{R}^{\nx}\}^T \mapsto \mathbb{R}$, there exists a $\sigma_h < \infty$ such that
\begin{align}
\sqrt{K}\left(\bar{h}_\text{\CPFAS}^{\Kmcmc}-\Exp{h(\s_{1:T})|\out_{1:T}}\right)
\end{align}
converges weakly to $\mathcal{N}(0,\sigma_h^2)$.
\end{thm}
\begin{pf}
The \CPFAS is uniformly ergodic for $N > 1$, \cite[Theorem 3]{lindsten2014particle}. Therefore \cite[Theorem 1.5.4]{Liang2010amc} is applicable.
\end{pf}

\note{The convergence of Algorithm~\ref{alg:mcmc_smoother} to the smoothing distribution is by Theorem~\ref{thm:conv_rate_cpfas} not dependent of the number of particles $N \rightarrow \infty$, but is \emph{only} relying on the number of iterations $K \rightarrow \infty$.

\subsection{Computational complexity}\label{sec:cpfas_cc}
The computational complexity of Algorithm~\ref{alg:mcmc_smoother} is of order $\mathcal{O}(KN)$, where $N$ is the number of particles in the \CPFAS and $K$ the number of iterations. However, in some programming languages, e.g., Matlab, vectorized implementations are preferable. The sequential nature of Algorithm~\ref{alg:mcmc_smoother} in $k$ does not allow such a vectorized implementation, which is a clear drawback. On the other hand, $K$ does not have to be specified a priori, but Algorithm~\ref{alg:mcmc_smoother} can be run repeatedly until satisfactory results are obtained, or a given computational time limit is violated.

The short message here is: The convergence rate for Algorithm~\ref{alg:mcmc_smoother} is $\sqrt{K}$, obtained at a computational cost of $\mathcal{O}(\Kmcmc)$ (for a fixed number of particles $N$). This can be compared to the convergence rate $\sqrt{N}$ to the less beneficial cost of $\mathcal{O}(N^2)$ for \FFBSi. However, one should remember that the samples obtained from \FFBSi are uncorrelated, which is typically not the case for Algorithm~\ref{alg:mcmc_smoother}.

\section{Simulated examples}
\subsection{Scalar linear Gaussian SSM}

As a first example, consider the scalar linear Gaussian SSM
\vspace{-1em}
\begin{subequations}\label{eq:LGSS}
\begin{align}
  \label{eq:LGSSa}
  x_{t+1} &= 0.2x_t + u_t + w_t, \quad &w_t& \sim \N(0,0.3),\\
  \label{eq:LGSSb}
  y_t &= x_t + e_t, \quad &e_t& \sim \N(0,1),
\end{align}
with $\Exp{x_1} = 0$ and $\Exp{x_1^2} = 0.1$. Implementing Algorithm~\ref{alg:mcmc_smoother} with $N = 2$ (with $T = 80$ and $u_t$ being low-pass filtered white noise), the result
shown in Figure~\ref{fig:intro} is obtained. As the system is linear and Gaussian, analytical expressions for $p(x_{1:T}|y_{1:T})$ can be found using the RTS smoother, shown in gray in Figure~\ref{fig:intro}.
\end{subequations}

\subsection{Nonlinear, multi-modal example}\label{sec:ex}

We will now turn to a more challenging problem, pinpointing some interesting differences between the forward-backward smoother (\FFBSi) and our MCMC-based\linebreak smoother in Algorithm~\ref{alg:mcmc_smoother}. We will start with a discussion using intuitive arguments, to motivate the example.

The \FFBSi smoother handles the path degeneracy problem in the particle filter discussed in Section~\ref{sec:particlemethods}. However, the support for the backward simulation is still limited to the particles sampled by the particle filter. As those particles, for $t < T$, are sampled from the \emph{filtering} distribution $p(\s_{t}|\out_{1:t})$ (and not the smoothing distribution $p(\s_{t}|\out_{1:T})$, due to the factorization \eqref{eq:ffilteq}), only few of the particles may be useful if the difference between the filtering and smoothing distribution is `large'. This might cause a problem for the \FFBSi smoother, since there might exist cases where the particles do not explore the relevant part of the state space. An interesting question is now if Algorithm~\ref{alg:mcmc_smoother} can be expected to explore the relevant part of the state space better than the \FFBSi smoother?

One way to understand the effect of the conditional trajectory in \CPFAS is as follows: If a proposal distribution $\proposal \neq f$ is used in a regular particle filter (Step~5 in Algorithm~\ref{alg:pf}), it is compensated for in the update of the weights, Step~7 and \eqref{eq:W}, so that $\{\s_{1:t}^i,\w_t^i\}_{i=1}^N$ are still an approximation of $p(\s_{1:t}|\out_{1:t})$, even if $\proposal \neq f$.

The \CPFAS can be thought of as a regular particle filter, but with a `proposal' $\proposal(x_t)$ that deterministically sets $\s_t^N = \s_t[\kmcmc]$ (Step~7 of Algorithm~\ref{alg:cpfas}) and `artificially' assigns an ancestor to it (Step~8). However, there is no compensation for this `proposal' in Step~10. Therefore, the samples $\{\s_{1:t}^i,\w_t^i\}_{i=1}^N$ from the \CPFAS can be expected to be biased towards the conditional trajectory $\s_{1:T}[k]$.

On the other hand, we know from \cite{lindsten2014particle} that the conditional trajectories in the limit $\kmcmc \rightarrow \infty$ are samples of $p(\s_{1:T}|y_{1:T})$. The bias towards $\s_{1:T}[\kmcmc]$ in the \CPFAS can therefore be thought of as `forcing' the \CPFAS to explore areas of the state space relevant for the smoothing distribution $p(\s_{1:T}|\out_{1:T})$ (rather than the filtering distribution $p(\s_{1:t}|\out_{1:t})$) for large $k$.

A simulated example,
appealing to this discussion, is now given. The problem is to sample from the smoothing distribution for a one-dimensional SSM with multi-modal properties of $g(\s_t|y_t)$. The state space model is $f(\s_{t+1}|\s_t) = \mathcal{N}(\s_{t+1}|\s_t,\sigma^2)$ and $g(\out_t|\s_t)$ is implicitly defined through the surface in Figure~\ref{fig:ex_1_res}, where the surface level in point $(\s,t)$ defines $g(\out_t|\s_t)$, for a given $\out_t$ (not shown).

Given $\s_0$, finding the maximum a posteriori estimate of the smoothing distribution $p(\s_{1:T}|\out_{1:T})$ amounts to finding the path $\s_{1:T}$ maximizing $p(\s_{1:T}|\out_{1:T}) \propto$ \\$\prod_{t=1}^T f(\s_{t}|\s_{t-1}) \prod_{t=1}^T g(\out_t|\s_t)$, where $g(\out_t|\s_t)$ is defined through the surface in Figure~\ref{fig:ex_1_res}. Intuitively, this can be thought of as going from left ($t=0$) to right ($t=100$) in Figure~\ref{fig:ex_1_res}, playing the children's game `the floor is hot lava' with the cost $f$ for moving sideways.

The mean of the filtering distribution $p(\s_{t}|\out_{1:t})$ for $t = 1, \dots, T$ (obtained by Algorithm~\ref{alg:pf}) is shown in Figure~\ref{fig:ex_1_res}, together with the mean from two different smoothers;  \FFBSi (Section~\ref{sec:ffbsi}) and Algorithm~\ref{alg:mcmc_smoother}, respectively.

\begin{figure}[t]
  \begin{center}
    \includegraphics[width=0.5\textwidth]{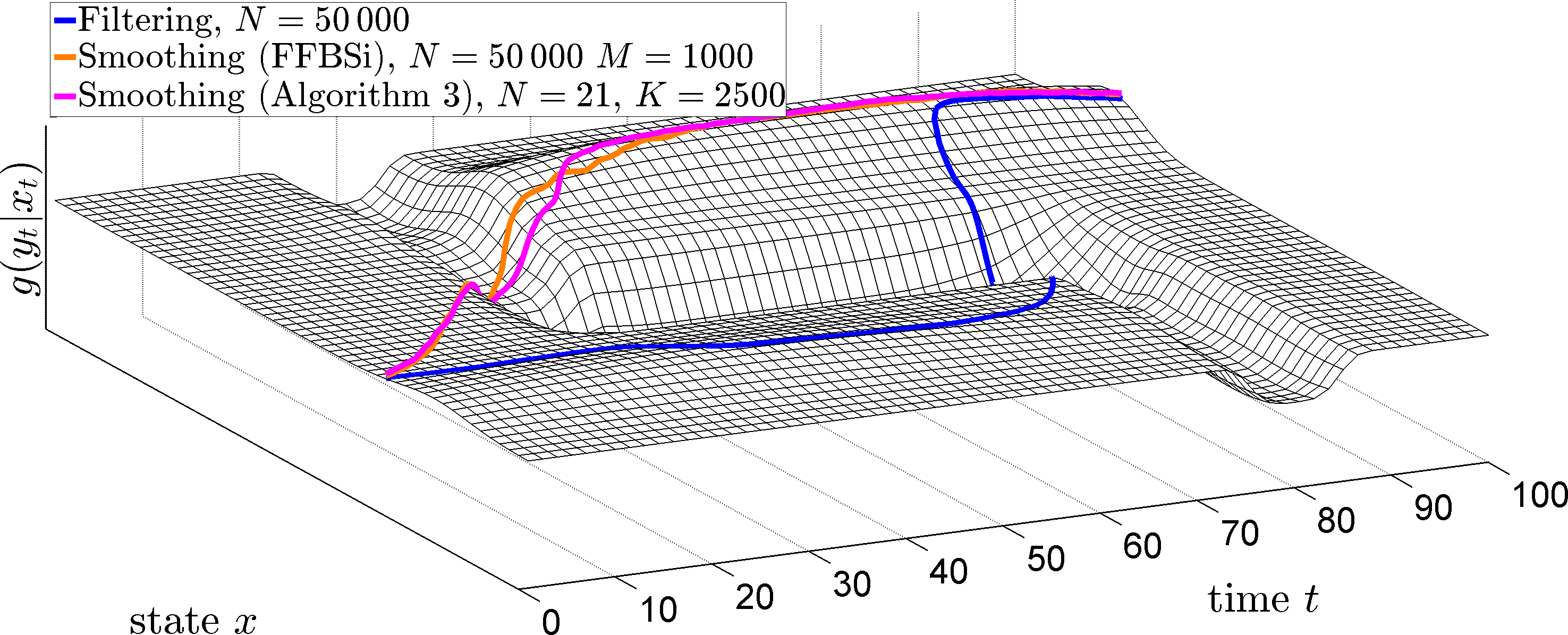}
  \end{center}
  \caption{\small The `landscape' for the Example in \ref{sec:ex}. The surface is proportional to $g(\out_t|\s_t)$, and the axis are time~$t$ and state~$x$, respectively. The trajectories are the mean of the results of a particle filter (Algorithm~\ref{alg:pf}, filtering), \FFBSi (smoothing) and Algorithm~\ref{alg:mcmc_smoother} (smoothing). Note that for both smoothing methods, a total of $50\thinspace000$ particles were sampled, so the comparison is `fair' in that sense.}
  \label{fig:ex_1_res}
\end{figure}

The two smoothing approximations can indeed be expected to approach each other in the limit $N \rightarrow \infty$ / $K \rightarrow \infty$. The problem is interesting because the `likelihood landscape' in Figure~\ref{fig:ex_1_res} contains a `trap'. The filtering distribution (and hence the particle filter in \ffbsi) will follow the right `shoulder' and discover `too late' (the valley at $t \approx 70$) that it `should' have walked along the left. The smoothing distribution, however, walks along the left shoulder earlier, as it `knows' that the valley at the right hand side will come.

To quantify this discussion on how well the particles explore the state space for the two smoothers, the densities of the sampled particles for both smoothers are plotted in Figure~\ref{fig:ex_1_dens}. This suggest that Algorithm~\ref{alg:mcmc_smoother} is able to give a better approximation of the smoothing distribution, as a larger proportion of the particles are sampled in a relevant part of the state space.

\begin{figure}[t]
  \begin{center}
    \includegraphics[width=0.48\textwidth]{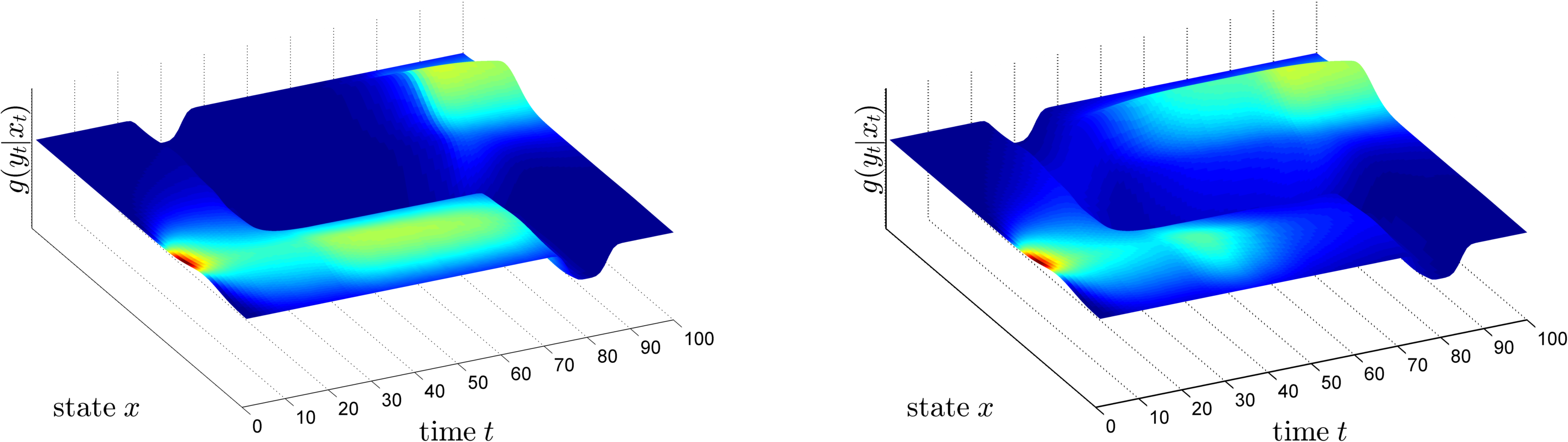}
  \end{center}
  \caption{\small Particle densities (on a blue-yellow-red scale, from low to high) for \FFBSi (left) and Algorithm~\ref{alg:mcmc_smoother} (right). In both cases, 50\thinspace000 particles are sampled for each $t$. They are, however, centered along the \emph{filtering} distribution for the \FFBSi, but biased toward the \emph{smoothing} distribution for Algorithm~\ref{alg:mcmc_smoother}. That is, Algorithm~\ref{alg:mcmc_smoother} explores (at least in this example) the relevant part of the state space (i.e., the left `shoulder') better.}
  \label{fig:ex_1_dens}
\end{figure}

\section{Indoor positioning application}

In this section, the presented algorithm is applied to a real-world sensor fusion problem; indoor positioning using ultrawideband (UWB), gyroscope and accelerometer measurements. We apply the model from \cite{kok2014indoor}, but rather than using the optimization-based approach in that paper, we employ Algorithm~\ref{alg:mcmc_smoother}. Instead of obtaining the Maximum a Posteriori (MAP) estimate as a point (as in \cite{kok2014indoor}), we will obtain samples from the posterior distribution, which can be used to estimate the MAP, mean, credibility intervals, etc.

\subsection{Problem setup}
We take the problem as presented by \cite{kok2014indoor}, a 10-dimensional nonlinear non-Gaussian problem. The goal is to estimate the position, velocity and orientation of the sensor board with the UWB transmitter, accelerometer and gyroscope, placed on the foot of a human. The UWB transmitter sends out pulses at (unknown) times $\tau_t$, and the time of arrival at the 10 receivers (indexed by $m$) are measured. The setup is calibrated using the algorithm in \cite{kok2014indoor}, making sure that the receiver positions $r_m^n$ are known and that their clocks are synchronized.

In the model, the state vector is $\s_t^T = [p_t^T v_t^T q_t^T]$, $p_t$ is the (3D) position, $v_t$ the velocity and $q_t$ the orientation (parametrized using unit quaternions). The \SSM is given by
\begin{subequations}
	\vspace{-0.5em}
\label{eq:ss_positioning}
	\begin{align}
	p_{t+1}^n &= p_t^n + T_sv_{t}^n + \tfrac{T_s^2}{2}a_t^n,\label{eq:eq}\\
	v_{t+1}^n &= v_t^n + T_sa_t^n, \\
	q_{t+1}^{nb} &= q_t^{nb} \odot \exp{\tfrac{T_s}{2}\omega_t},\label{eq:eq3}\\
	y_{m,t} &= \tau_{t} + \tfrac{1}{c}\| r_m^n-p_t^n \|_2 + e_{m,t} \label{eq:eq4}
	\end{align}
\end{subequations}
where \eqref{eq:eq}~--~\eqref{eq:eq3} are the dynamics and \eqref{eq:eq4} is the measurement equation. The superscripts $n$ and $b$ denote coordinate frames ($n$ is the navigation frame aligned with gravity, and $b$ is the body frame, aligned with the sensor axes of the accelerometers), $c$ denotes the speed of light, $\odot$ denotes the quaternion product and $\exp$ denotes the vector exponential \citep{Hol:2011}. $T_s$ is the time between two data samples from the accelerometer and gyroscope, sampled with $120$\thinspace Hz. However, the UWB samples are sampled at approximately $10$\thinspace Hz. Due to the nature of UWB measurements, $e_{m,t}$ is modeled as
\begin{align}
e_{m,t} \sim  \begin{cases}
    (2-\alpha)\mathcal{N}(0,\sigma^2), & e_{m,t} < 0,\\
    \alpha\text{Cauchy}(0,\gamma), & e_{m,t} \geq 0,
  \end{cases}
\end{align}
because measurements can only arrive later (and not earlier) in case of multipath and non-line-of-sight propagation. 
The acceleration $a_t^n$ is found via accelerometer measurements $y_{a,t}$, modeled as
\begin{align}
	y_{a,t} = R_t^{bn}(a_t^n-g^n) + \delta_a + e_{a,t},
\end{align}	
with $g^n$ denoting gravity, $R_t^{nb}$ is a rotation matrix representation of $q_t^{nb}$, and $R_t^{bn} = (R_t^{nb})^T$. The angular velocity $\omega_t$ is obtained from the gyroscope measurements $y_{\omega,t}$ as
\begin{align}
	y_{\omega,t} = \omega_t + \delta_{\omega} + e_{\omega,t}.\label{eq:omega}
\end{align}
The noise $e_{a,t}$ and $e_{\omega,t}$ are modeled as $\mathcal{N}(0,\sigma_a^2)$ and $\mathcal{N}(0,\sigma_\omega^2)$, respectively. $\delta_a$ and $ \delta_{\omega}$ are sensor biases. Note that the accelerometer and gyroscope measurements are not treated as outputs in \eqref{eq:ss_positioning}, but rather as inputs to the dynamics, implicitly introducing an uncertainty in $f(\s_{t+1}|\s_t)$ through the measurement noise.

\subsection{Results}

\begin{figure}[b]
  \begin{center}
	\begin{subfigure}[b]{0.95\columnwidth}
		\includegraphics[width=0.95\columnwidth]{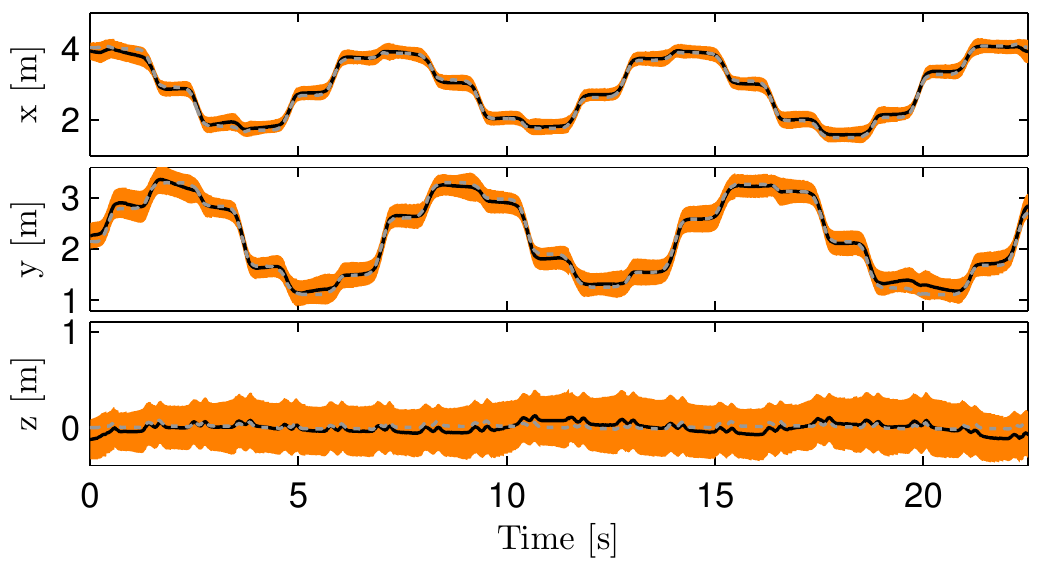}
		\caption{Smoothing distribution for position.}
		\label{fig:positions}
	\end{subfigure}
	~
	\begin{subfigure}[b]{0.95\columnwidth}
		\includegraphics[width=0.95\columnwidth]{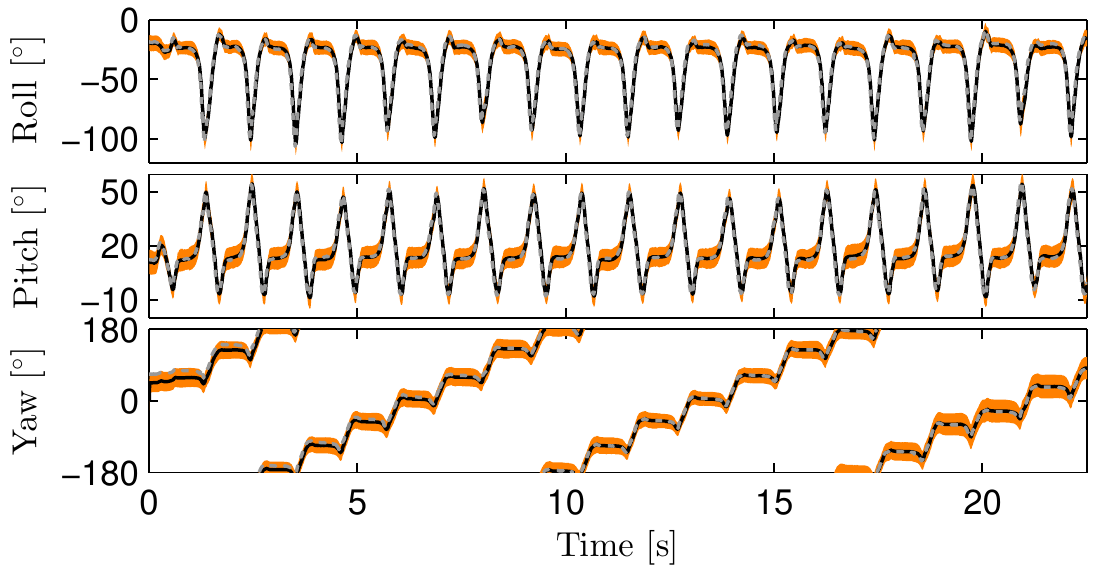}
		\caption{Smoothing distribution for orientation.}
		\label{fig:angles}
	\end{subfigure}
	\caption{\small The mean (black line) and 99\% credibility intervals (orange fields) of $p(\s_{1:T}|\out_{1:T})$ for the position and orientation states, and ground truth (dashed gray) from an optical reference system.}
	\label{fig:pos_res}
	\end{center}
\end{figure}

Algorithm~\ref{alg:mcmc_smoother} was applied to data presented by \cite{kok2014indoor} with \eqref{eq:ss_positioning}~--~\eqref{eq:omega}. The results for $K = 1000$ iterations and $N = 500$ particles are summarized in Figure~\ref{fig:pos_res} in terms of the mean and credibility intervals (cf. Figure~13 and 14 in \cite{kok2014indoor}). For reference, the ground truth (obtained by an optical reference system) is also shown in the plot. In terms of computational load, the presented results took about 1 day to obtain on a standard desktop computer. 

Note the credibility intervals, which are the gain of using this method producing samples (as opposed to a method based on point estimates). The credibility intervals are varying over time and are different for different states, which indeed adds information to the results.

\section{Conclusions}

We have shown how the CPF-AS can be used to solve the nonlinear state smoothing problem in a disparate way compared to the currently available particle smoothers. The asymptotic convergence of our smoother was established, and we also illustrated the use of the smoother on two simulated examples and one challenging real-world application.

Based on the results of Theorem~\ref{thm:conv_rate_cpfas} and the numerical examples we conclude that Algorithm~\ref{alg:mcmc_smoother} is indeed a competitive alternative to the existing state-of-the-art smoothers. The present development opens up for interesting future work, such as hybrid versions of \ffbsi and Algorithm~\ref{alg:mcmc_smoother}, where \ffbsi is used to initialize Algorithm~\ref{alg:mcmc_smoother}. Further studies on how to tackle the trade-off between the number of particles $N$ and the number of iterations $K$ in Algorithm~\ref{alg:mcmc_smoother} for optimal performance (given a computational limit) would also be interesting.

\footnotesize
\section*{Acknowledgments}
\vspace{-1em}
We would like to thank Dr. Jeroen Hol and Dr. Henk Luinge at Xsens Technologies for providing the indoor positioning data.

\bibliographystyle{ifacconf-harvard}
\bibliography{cpfas_refs}

\newpage


\section*{Supplementary material}

# Some details on state space smoothing using the conditional particle filter[⋆]

Andreas Svensson[∗] Thomas B. Schön[∗] Manon Kok[∗∗]

[∗] *Department of Information Technology, Uppsala University, Sweden*
*(e-mail: {andreas.svensson, thomas.schon}@it.uu.se)*
[∗∗] *Division of Automatic Control, Linköping University, Sweden*
*(e-mail: manon.kok@liu.se)*

**Abstract:** This technical report gives some additional details on the numerical examples used in Svensson et al. (2015).

## 1. INTRODUCTION

This report contains some details on the examples presented by Svensson et al. (2015). The problem under consideration is that of state space smoothing using the conditional particle filter with ancestor sampling (CPF-AS). In this report, we will address additional details concerning

i) the simulated nonlinear, multimodal example used in Section 4.2.
ii) the indoor positioning application used in Section 5.

Matlab code for the simulated example can be found via the first author's homepage.

## 2. SIMULATED NONLINEAR, MULTIMODAL EXAMPLE

This section gives an extended introduction to the problem considered in Svensson et al. (2015, Section 4.2). The essential details are contained in the original article, but we will provide an alternative introduction. The state space model considered in the problem is

$$x_{t+1}|x_t \sim \mathcal{N}(x_t, 0.5), \quad (1a)$$
$$y_t|x_t \sim g(y_t|x_t), \quad (1b)$$
$$x_1 = 3. \quad (1c)$$

The problem concerns smoothing for a given output sequence $y_{1:t}$. Thus we write $G(x_t) \triangleq g(y_t|x_t)$, the likelihood of $x_t$, omitting the dependence on $y_t$. $G(x_t)$ is defined by the surface of Figure 1, which is parametrized as a sum of six Gaussian-like functions. The specific parametrization is found in the provided source code.

The state space filtering/smoothing problem arising from this problem is solved using three approaches. For filtering, the particle filter in Svensson et al. (2015, Algorithm 1), with $q_t = f$, is used. For smoothing, the FFBSi with rejection sampling, as presented in Lindsten and Schön (2013, Algorithm 5), is used, together with the MCMC smoother in Svensson et al. (2015, Algorithm 3), and the results reported by Svensson et al. (2015) are obtained.

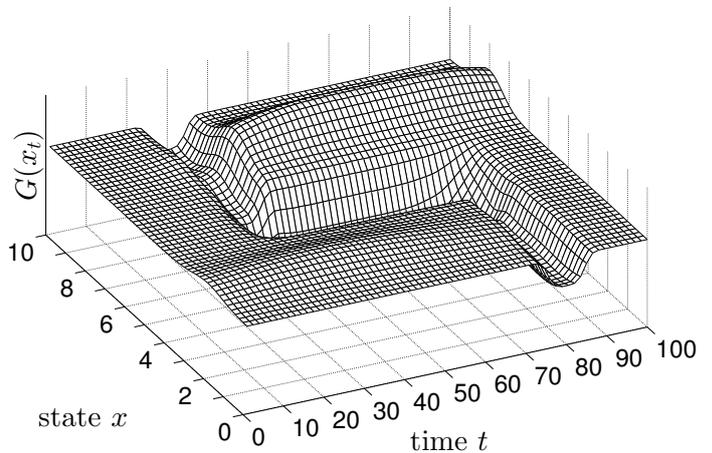

Fig. 1. Definition of $G(x_t)$.

## 3. INDOOR POSITIONING

This section focuses on the indoor positioning application in Svensson et al. (2015, Section 5). The problem is discussed in Kok et al. (2015), and concerns a sensor fusion problem involving accelerometer, gyroscope, and ultrawideband (UWB) measurements. There are several details on the problem not covered by Svensson et al. (2015); the non-uniform sampling intervals of the UWB measurements, the unknown transmission times, the sensor biases, some issues on the ancestor sampling, and the initialization of the first conditional trajectory.

We repeat the state space model here:

$$\left.\begin{array}{l} p_{t+1}^n = p_t^n + T_s v_t^n + \frac{T_s^2}{2}a_t^n, \\ v_{t+1}^n = v_t^n + T_s a_t^n, \\ q_{t+1}^{nb} = q_t^{nb} \odot \exp\frac{T_s}{2}\omega_t, \end{array}\right\} f(x_{t+1}|x_t, a_t^n, \omega_t) \quad (2a)$$

$$y_{m,t} = \tau_t + \|r_m^n - p_t^n\|_2 + e_{m,t} \quad \} g(y_t|x_t), \quad (2b)$$

with the same notation as Svensson et al. (2015). Most important, $x_t^{\mathrm{T}} = [(p_t^n)^{\mathrm{T}} \ (v_t^n)^{\mathrm{T}} \ (q_t^{nb})^{\mathrm{T}}]$ forms the state vector, and $a_t^n$ are the acceleration, $\omega_t$ the angular velocity, and $y_{m,t}$ are the UWB measurements (for receiver number $m$). The sampling frequency for the accelerometers and gyroscopes is 120 Hz.

The acceleration and angular velocity are modeled to be measured as

$$y_{a,t} = R_t^{bn}(a_t^n - g^n) + \delta_a + e_{a,t}, \quad (3)$$
$$y_{\omega,t} = \omega_t + \delta_\omega + e_{\omega,t}. \quad (4)$$

with notation as Svensson et al. (2015). Through the measurement noise in these models, stochastic noise enters the state space.

### 3.1 Non-uniform sampling interval

For every sampling step $t$, the accelerometer and gyroscope data $y_{a,t}$ and $y_{\omega,t}$ are available. However, the UWB measurements are recorded at a lower sampling rate, and are available only for approximately every tenth sample. A high level algorithm illustrating how this is handled is shown in Algorithm 1.

**Algorithm 1** Algorithmic strategy for the indoor positioning problem

1: **for** $k = 1, \ldots, K$ **do**
2:   Sample $x_1^i$ from $q_1(x_1)$
3:   Draw $x_1^i \sim p(x_1^i), i \in [1, N-1]$.
4:   Set $x_{1:T}^N = x_{1:T}[k]$.
5:   **for** $t = 1, \ldots, T-1$ **do**
6:     **if** UWB measurement $\{y_{m,t}\}_{m=1}^M$ available **then**
7:       Set $w_t^i = g(y_t|x_t^i), i \in [1, N]$.
8:       Draw $b_t^i$ with $\mathbb{P}\left(b_t^i = j\right) \propto w_{t-1}^j, i \in [1, N-1]$.
9:       Set $x_{1:t}^i = \{x_{1:t-1}^{b_t^i}, x_t^i\}, i \in [1, N-1]$.
10:     **end if**
11:     Draw $x_{t+1}^i$ from $f(x_{t+1}^i|x_t^i, a_t, \omega_t), i \in [1, N-1]$.
12:     Draw $b_{t+1}^N$ with $\mathbb{P}\left(b_{t+1}^N = j\right) \propto f(x_{t+1}^N|x_t^j)$.
13:     Set $x_{1:t+1}^N = \{x_{1:t}^{b_{t+1}^N}, x_{t+1}^N\}$.
14:   **end for**
15:   Draw $J$ with $\mathbb{P}(i = J) \propto \frac{1}{N}$ and set $x_{1:T}[k+1] = x_{1:T}^J$.
16: **end for**

*(Note that the notation for the ancestor sampling variable is $b$, instead of $a$, not to confuse it with the accelerometer data.)*

### 3.2 Unknown transmission times

When evaluating (2b), the transmission time $\tau_t$ is unknown. It is approximately handled by Monte Carlo integration:

$$g(y_t|x_t) = p_e(y_{m,t} - \tau_t - \|r_m^n - p_t^n\|_2 \mid \tau_t)$$
$$\tilde{\propto} \sum_j p_e(y_{m,t} - \tau_t^j - \|r_m^n - p_t^n\|_2), \quad (5)$$

where $p_e$ is the pdf defined by eq. (11) in Svensson et al. (2015), and $\tau_t^j$ are samples with possible values of $\tau$. All particles (indexed with $i$) are evaluated with the same set of samples $\tau_t^j$.

### 3.3 Sensor bias

The numerical values of the sensor biases $\delta_a$ and $\delta_\omega$ are small, compared to the noise levels. They were therefore approximated manually. However, a more thorough system identification approach can be applied within the CPF-AS framework, as proposed by Lindsten (2013).

### 3.4 Evaluation of $f(x_{t+1}|x_t, a_t^n, \omega_t)$

For the ancestor sampling in Step 1 in Algorithm 1, $f(x_{t+1}|x_t, a_t^n, \omega_t)$ needs to be evaluated. This involves the quaternion product and vector exponential in (2a), which can be manipulated as follows:

$$q_{t+1}^{nb} = q_t^{nb} \odot \exp \tfrac{T_s}{2}(\omega_t + e_{\omega,t}) \Leftrightarrow$$
$$e_{\omega,t} = \tfrac{2}{T_s} \log\left((q_t^{nb})^{-1} \odot q_{t+1}^{nb}\right) - \omega_t \quad (6)$$

where log denotes the logarithm for unit quaternions (Hol, 2011). An approximation of this expression was used in the computations.

### 3.5 Low chance of 'new ancestor'

One challenge is the small uncertainty in the position (2a), because of the (physically reasonable) factor $\frac{T_s^2}{2}$ ($\approx 10^{-5}$) in front of the noise term. To handle this, the uncertainty was artificially increased in the ancestor sampling step. This causes a substantial increase in the mixing, but also a non-feasible 'jump' in the smoothing trajectories. These artificial 'jumps' appears, however, to be rare and can also be expected to 'even out' as $K \to \infty$, but might indeed cause an overestimate of the variance.

A more thorough treatment would be to apply the recent development by Lindsten et al. (2015).

### 3.6 Initialization

The model has a state space of 9 dimensions (parametrized using 10 variables). The structure of the model and the uncertainties cause, according to our experience, the particle filter to diverge quickly if $N$ is not sufficiently large. To speed up the initialization phase, a more 'loose' model was used in the first run of the CPF-AS to find one reasonable trajectory. This non-diverging trajectory was then used as the conditional trajectory for the subsequent run with the correct model.


\end{document}